\documentstyle[12pt,aaspp4]{article}
\def\hst{{\sl HST}}
\def\h0{H$_0$}
\def\etal{et al.\ }
\def\magdex{mag dex$^{-1}$}
\begin{document}
\onecolumn
\title{The {\sl HST} Key Project on the Extragalactic Distance Scale XIII. \\
The Metallicity Dependence of the Cepheid Distance Scale 
\footnote{Based on observations with the NASA/ESA
{\it Hubble Space Telescope} obtained at the Space Telescope Science
Institute, which is operated by AURA, Inc. under NASA Contract No.
NAS5-26555.} 
\footnote{Observations reported here have been obtained in part with 
the Muliple Mirror Telescope (MMT), a joint facility of the University
of Arizona and the Smithsonian Institution.}}

\author{Robert C. Kennicutt, Jr.,\footnote{Steward Observatory,
University of Arizona, Tucson, AZ 85721}
Peter B. Stetson,\footnote{Dominion Astrophysical Observatory,
Victoria, British Columbia V8X 4M6 Canada}
Abhijit Saha,\footnote{Space Telescope Science Institute,
3700 San Martin Drive, Baltimore, MD 21218}
Dan Kelson,\footnote{Lick Observatory, University of California, 
Santa Cruz, CA  95064}
Daya M. Rawson,\footnote{Mount Stromlo and Siding Spring
Observatories, Institute of Advanced Studies, ANU, ACT 2611 Australia}
Shoko Sakai,\footnote{Infrared Processing and Analysis Center,
California Institute of Technology, Pasadena, CA  91125}
\addtocounter{footnote}{-1}
Barry F. Madore,\footnotemark
\addtocounter{footnote}{-2}
Jeremy R. Mould,\footnotemark
\addtocounter{footnote}{+1}
Wendy L. Freedman,\footnote{Observatories of the Carnegie
Institution of Washington, Pasadena CA 91101}
\addtocounter{footnote}{-7}
Fabio Bresolin,\footnotemark
\addtocounter{footnote}{+6}
Laura Ferrarese,\footnote{Department of Astronomy, California Institute
of Technology, Pasadena, CA 91125}
Holland Ford,\footnote{Department of Physics and Astronomy, Johns Hopkins
University, Baltimore, MD 21218}
\addtocounter{footnote}{-5}
Brad K. Gibson,\footnotemark
\addtocounter{footnote}{+4}
John A. Graham,\footnote{Department of Terrestrial Magnetism,
Carnegie Institution of Washington, Washington D.C. 20015}
Mingsheng Han,\footnote{Department of Astronomy, University of Wisconsin,
475 North Charter Street, Madison, Wisconsin 53706}
\addtocounter{footnote}{-11}
Paul Harding,\footnotemark
\addtocounter{footnote}{+9}
John G. Hoessel,\footnotemark
John P. Huchra,\footnote{Harvard Smithsonian
Center for Astrophysics, 60 Garden Street, Cambridge, MA 02138}
Shaun M.G. Hughes,\footnote{Royal Greenwich
Observatory, Cambridge CB3 0EZ England, UK}
\addtocounter{footnote}{-10}
Garth D. Illingworth,\footnotemark
\addtocounter{footnote}{+7}
Lucas M. Macri,\footnotemark
\addtocounter{footnote}{-6}
Randy L. Phelps,\footnotemark 
\addtocounter{footnote}{-2}
Nancy A. Silbermann,\footnotemark
\addtocounter{footnote}{-6}
Anne M. Turner,\footnotemark
\,and
\addtocounter{footnote}{+3}
Peter R. Wood\footnotemark
\addtocounter{footnote}{+8}
}

\newpage

\begin{abstract}

Uncertainty in the metal abundance dependence of the Cepheid 
variable period-luminosity (PL) relation remains one of the outstanding 
sources of systematic error in the extragalactic distance scale and the 
Hubble constant. To test for such a metallicity dependence, we have 
used the Wide Field Planetary Camera 2 (WFPC2) 
on the {\sl Hubble Space Telescope} 
(\hst) to observe Cepheids in two fields in the nearby spiral galaxy M101, 
which span a range in oxygen abundance of 0.7 $\pm$ 0.15 dex.  A 
differential analysis of the PL relations in $V$ and $I$ in the two fields 
yields a marginally significant change in the inferred distance modulus on 
metal abundance, with  $\delta(m - M)_0/\delta[O/H] = -0.24 \pm 0.16$ 
\magdex.  The trend is in the theoretically predicted sense that 
metal-rich Cepheids appear brighter and closer than metal-poor stars. 
External comparisions of Cepheid distances with those derived from three 
other distance indicators, in particular the tip of the red giant
branch method, further constrain the magnitude of any  
$Z$-dependence of the PL relation at $V$ and $I$. 
The overall effects of any 
metallicity dependence on the distance scale derived with \hst\ will be of 
the order of a few percent or less for most applications, though 
distances to individual galaxies at the extremes of the metal abundance 
range may be affected at the 10\% level.

\end{abstract}


\section{Introduction}

Cepheid variable stars play a central role in the calibration of the 
extragalactic distance scale and the Hubble constant (\h0).  
The measurement of Cepheid 
distances to a dozen galaxies with the \hst\ has led to dramatic 
improvements in the calibration of the several secondary distance 
indicators,  and narrowed the longstanding discrepancy between the ``long" 
and ``short" distance scales from $40 - 100$ km~s$^{-1}$~Mpc$^{-1}$ a few 
years ago to $\sim55 - 75$ km~s$^{-1}$~Mpc$^{-1}$ currently (e.g., Saha 
\etal 1997; Madore \etal 1997).  The completion of the \hst\ Key Project 
on the Extragalactic Distance Scale will double again the number of 
galaxies with Cepheid distances, furnish firm calibrations of the 
Tully-Fisher (T-F), surface brightness fluctuation (SBF), planetary nebula 
(PNLF) and 
globular cluster luminosity functions, SN II expanding parallax, 
and SN Ia 
secondary distance indicators, and determine \h0\ to an accuracy of 
$\pm$10\% (Kennicutt, Freedman, \& Mould 1995).

As the uncertainty in the distance scale as a whole is reduced, systematic 
uncertainties in the Cepheid distances themselves become increasingly 
important, because errors in the Cepheid scale propagate with nearly full 
weight into the secondary distance ladder and \h0.  Of particular concern 
is whether the Cepheid period-luminosity (PL) relation varies 
systematically with metal abundance.  The metal abundances in the Key 
Project galaxy sample vary by nearly an order of magnitude (Zaritsky, 
Kennicutt, \& Huchra 1994; hereafter ZKH).  Current calibrations of the PL 
relation are referenced to the Large Magellanic Cloud, which has a lower 
metal abundance ([O/H] $\simeq -0.4$) than is often found in more 
luminous galaxies in the Key Project sample.  Thus a significant 
$Z$-dependence to the PL relation could introduce a systematic error in 
the overall distance scale, as well as an increased scatter in individual 
distances.  

Several theoretical and observational studies argue for a significant 
metallicity dependence, though the magnitude of the effect remains 
controversial.  Theoretical models by Chiosi, Wood, \& Capitanio (1993; 
hereafter CWC93) predict a small change in the PL zeropoint at visual and 
near-infrared wavelengths, of order 0.1 mag per factor of ten increase in 
abundance, but earlier calculations predicted a stronger effect (e.g., 
Stothers 1988), and the theoretical models remain uncertain (\S2).  

Evidence for larger effects have come from several recent empirical 
studies.  Freedman \& Madore (1990) undertook the first direct 
observational test of the metallicity dependence of the PL relation, using 
$BVRI$ photometry of Cepheids in three fields in M31.  
They derived a change in distance 
modulus of $-0.32 \pm 0.21$ \magdex, a difference which they did not 
regard as significant.  Their data were subsequently reanalyzed by Gould 
(1994), who derived steeper dependences of $-0.88 \pm 0.16$ and $-0.56 \pm 
0.20$ \magdex, depending on the bandpasses used in the fitting.  Recently 
Beaulieu \etal (1997) and Sasselov \etal (1997) analyzed EROS data for 
large samples of Cepheids in the LMC and SMC, and derived a change in 
inferred distance modulus  of $-0.44{^{+0.1}_{-0.2}}$ \magdex. Kochanek 
(1997) derived a similar dependence of $-0.4 \pm 0.2$ \magdex\ , 
using published Cepheid measurements 
of 17 galaxies.  Most of the dependence arises from increased line 
blanketing and redder intrinsic colors at higher abundance, an effect that 
has been reported in Galactic and Magellanic Cloud Cepheids by several 
authors (e.g., Caldwell \& Coulson 1986; Laney \& Stobie 1994, and 
references therein).  

All of these results are consistent with a metallicity dependence that 
tends to cause the distances of metal-rich Cepheids to be underestimated. 
However the magnitudes of these claimed dependences range over an order of 
magnitude, so the significance of any $Z$-dependence for the distance 
scale remains unclear.  A zeropoint change at the  level of $\le$0.3 
\magdex\ would have minimal effects on the ultimate distance 
scale, because the Cepheid target galaxies observed with \hst\ bracket a 
large metallicity range with a mean value close to that of the LMC 
Cepheids that calibrate the PL relation (\S6).  However 
dependences as large as those derived by Gould (1994) would introduce 
systematic errors in \h0\ at the 10--20\% level, and constitute the 
dominant systematic error in the entire extragalactic distance ladder.  
As pointed out by Sasselov et al (1997) and Kochanek (1997), a strong
metallicity dependence might account for part of the difference between
low values of \h0\ determined from SN~Ia and other secondary indicators,
if the Cepheid abundances in the SN~Ia hosts are systematically lower.
Such effects might also partly explain the longstanding difference in
Cepheid and RR Lyrae distance scales in nearby galaxies.  
It is imperative to test whether systematic errors at this level are present 
in the Cepheid data.

As part of the \hst\ Key Project we undertook a further test of the 
metallicity dependence of the PL relation, by targeting two fields in the 
giant Sc I spiral M101 (Kennicutt \etal 1995).  M101 is especially well 
suited for a metallicity test, because it combines an unusually steep 
radial abundance gradient with a large disk scale length, making it 
possible to identify Cepheids with \hst\ nearly to the center of the 
galaxy.  The fields we targeted differ in metal abundance by a factor of 
five, and bracket the abundance range between the LMC and the most 
metal-rich Cepheid fields observed with \hst.  In this paper we present a 
{\it differential} analysis of the Cepheid PL relations in the two fields.
Our approach differs from that taken in most of the previous studies, 
insofar as we direct our attention solely to the effects of abundance
on the $V$ and $I$ PL relations used in most of the \hst\ Cepheid observations.
We apply the same PL fitting methods and abundance calibrations 
that are applied to other Key Project observations, so that we can  
place direct limits on the effects of abundance variations on 
the \hst\ Cepheid distance scale, independent of any possible Cepheid
metallicity dependence at other wavelengths.  
We have also compared Cepheid distances 
to 23 galaxies, spanning a 50-fold range in abundance, with distances 
derived from three other methods, the tip of the red giant branch method 
(TRGB), the planetary nebula luminosity function (PNLF), and the 
Tully-Fisher (TF) method, to further constrain the magnitude of any 
metallicity dependence.  

The paper is organized as follows.  In \S2 we briefly summarize the 
effects of abundance on Cepheid magnitudes and colors as expected from 
theory, and present a series of simulations that quantify the effects on 
the $V$ and $I$ PL fits used in the \hst\ observations.  \S3 describes  
the differential test in M101 and the derived metallicity dependence. We 
compare the results to previous work in \S4.  In \S5  we use TRGB, PNLF, 
and TF distances to further constrain the metallicity dependence.  
Finally, in \S6 we assess the impact of the dependence on the overall 
distance scale, and discuss the prospects for a definitive calibration of 
the $Z$ dependence.

\section{Theoretical Expectations and Simulations}

The effects of metal abundance on the Cepheid instability strip have been 
investigated by Stothers (1988), Stift (1990; 1995), and CWC93. The main 
predictions are that metal-rich Cepheids at a given period are more 
luminous and redder.  The net effects on the observed PL relations are 
strongly wavelength dependent; metal-rich Cepheids will appear fainter in 
the blue due to line blanketing, while in the red the sense of the effect 
is to make the Cepheids appear brighter.  The magnitudes of the 
metallicity effects are also strongly wavelength dependent.  In this paper 
we are concerned solely with the behavior of the PL relations in $V$ and 
$I$, which is most relevant to \hst\ observations.

The methodology used to fit the PL relations observed in the Key Project 
is described in Freedman \etal (1994) and Ferrarese \etal (1996).  
Briefly, the Cepheid target fields are observed at 12 epochs in $V$ 
(F555W), and at $4 - 8$ epochs in $I$ (F814W).  The observed PL relations 
at $V$ and $I$ are fitted with slopes fixed to those of the calibrating PL 
relations in the LMC (Madore \& Freedman 1991):
\begin{equation}
M_V = -2.76 \log P - 1.40
\end{equation}
\begin{equation}
M_I = -3.06 \log P - 1.81
\end{equation}

These calibrations assume an LMC true distance modulus of 18.50~$\pm$~0.10 
and an average reddening $E(B-V) = 0.10$, corresponding to $E(V-I) = 
0.13$.  Any difference in the apparent $V$ and $I$ moduli is assumed to be 
caused by reddening, and the true modulus is computed assuming $A_V/A_I$ = 
1.666, or $A_V/E(V-I)$ = 2.50 (Cardelli, Clayton, \& Mathis 1989).
This is the same value found by Stanek (1996).

There are several ways in which metallicity effects could propagate into 
the Cepheid distances derived in this way.  Aside from directly 
influencing the absolute magnitudes of the Cepheids, any systematic change 
in $V-I$ color will propagate through the reddening correction and alter 
the inferred true modulus. CWC93 used analytical fits to their theoretical 
models to predict the magnitude of these effects for PL relations observed 
in $BV$ and $VI$. For a mean period of 10 days, their models predict 
$\delta (m-M)_0$ = $+$22.2$\delta Z$ from fitting PL relations in $B$ and 
$V$, but only $-$1.7$\delta Z$ when fitted in $V$ and $I$, where $Z$ is 
the metal mass fraction ($Z_\odot$ = 0.02).  The $VI$ dependence 
corresponds to a systematic error of only 0.06 mag in true modulus (3\% in 
distance) between $0.2 - 2~Z_\odot$, approximately the full range found in 
the Cepheid galaxies observed with \hst. Such a weak dependence would be 
virtually impossible to measure in the \hst\ data.  The predicted effects 
on distances derived from photometry in $B$ and $V$ would be much larger, 
44\% (or 0.8 mag) for the same abundance range.

However there are other, more subtle ways in which abundance effects might 
propagate, for example in the way that the width of the Cepheid 
instability strip and the colors of the blue and red edges are affected 
(CWC93). In order to incorporate these effects, we constructed simulated 
Cepheid data sets for different metal abundances and foreground 
reddenings, and analyzed them using the standard Key Project reduction 
procedures.  Simulated PL relations for a given metallicity $Z$ were 
derived using the equations given in CWC93. Cepheid masses were then 
randomly drawn from a mass distribution given by $dn/dm = m^{-3}$ over the 
mass range $3 - 13~M_\odot$. Luminosities were calculated from equation 
(25) of CWC93, and the red and blue edges of the instability strip were 
derived using equations (7) and (13) of the same paper.  Stars were 
randomly positioned between the red and blue edges of the strip, and 
periods were then calculated from the P--M--L--T$_{eff}$ relation given in 
equation (4) of CWC93. We used fits to the ($BC_V$,$T_{eff}$) and ($V-I$,
$T_{eff}$) relations to  derive the apparent $V$ and $I$ magnitudes. 
Reddening was then added assuming $E(V-I) = 1.31 E(B-V)$ and $A(V) = 3.1 
E(B-V)$.\footnote{These reddening coefficients differ slightly from those 
used in our standard data reductions, but the difference is not 
significant for this application.} Nine samples of 100 Cepheids each were 
constructed, with metal abundances $Z$ = 0.01, 0.02, and 0.03, and 
reddening $E(B-V)$ = 0.1 and 0.2 mag.  The simulated data sets
were then fitted to PL relations in $V$ and $I$, following the 
standard procedures described earlier.  The results are summarized in 
Table 1.  The first three columns of the table are the input values to the 
simulation, while the last four columns are derived quantities.

We then determined the coefficient in equation (3) by minimizing
the difference between $(m - M)_{true}$ and $(m - M)_{PL}$.
\begin{equation}
(m - M)_{true} = (m - M)_{PL} + (0.11 \pm 0.03)~\log {(Z/Z_{LMC})}
\end{equation}

\noindent
Throughout this paper we will use the notation $\gamma$ to denote the 
change in inferred true distance modulus per factor of ten in metal 
abundance (cf. Beaulieu \etal 1997; Kochanek 1997), with a subscript
indicating the relevant wavelengths over which the PL relation is 
fitted ($V$ and $I$ unless indicated otherwise). Thus for our 
simulations $ \gamma_{VI} = -0.11 \pm 0.03$~\magdex.  

The metallicity effect derived in our simulations is slightly larger than 
the analytical approximation of CWC93, but is still very slight, amounting 
to a distance error of less than 6\% for an order of magnitude change in 
metal abundance. The sign of the change is in the sense that the distance 
inferred at higher metal abundances is smaller (Cepheids appear brighter). 
Unfortunately the theoretical prediction is not robust.  For example, the 
theoretical prediction of the position of the red edge of the instability 
strip is poorly defined by theory, because of large uncertainties in the 
theory of convective transport.  More recent calculations by Chiosi, Wood, 
\& Capitanio (1997), using new opacities, predict $\gamma_{VI}$ = $+$0.06 
\magdex, a smaller effect but one in the opposite sense.  Although it is 
reassuring that the metallicity dependence predicted by theory is small, 
we cannot yet use theory as the basis of a correction to observed distance 
moduli.  Moreover it would be very useful to have independent 
observational confirmation of the magnitude of the metallicity effects, if 
not an empirical calibration of the dependence.

\section{A Differential Test in M101}

The magnitude of the $Z$-dependence of the PL relation can be measured 
directly, by observing Cepheids with different metal abundances in the 
same galaxy.  This approach was introduced by Freedman \& Madore (1990), 
who applied the method to M31 using groundbased $BVRI$ CCD photometry.  As 
part of the \hst\ Key Project, we observed Cepheids in two fields in the 
giant Sc galaxy M101.  The ``inner" and ``outer" fields are centered at  
14$^{\rm h}$03$^{\rm m}$23$^{\rm s}$.94, $+$54\arcdeg21\arcmin35\arcsec.7 
and 14$^{\rm h}$02$^{\rm m}$22$^{\rm s}$.49, 
$+$54\arcdeg17\arcmin58\arcsec.3 (2000.0), at galactocentric radii of 
1\arcmin.7 and 7\arcmin.9, respectively.  Figure 1 shows a groundbased
image of M101 with the inner and outer WFPC2 field locations superimposed.

\subsection{Abundances}

The disk abundances in M101 are among the best studied of any galaxy, 
thanks to several HII region surveys (Kennicutt \& Garnett 1996 and 
references therein).  The HII region abundances are measured in terms of 
oxygen, and we will parametrize the Cepheid metallicity dependence 
in terms of the nebular oxygen scale using the usual notation: [O/H] 
$\equiv \log(O/H)/(O/H)_\odot$, and adopting 
$(O/H)_\odot = 7.9 \times 10^{-4}$.  
We adopt this convention because metal abundances for most of the Key 
Project Cepheid fields have been measured from HII region spectra (ZKH).  
The abundances of the Cepheids themselves could differ slightly from those 
of the HII regions, either because the stellar iron and oxygen abundances 
do not scale precisely, or from a small calibration offset between the stellar 
and interstellar abundances.  Neither of these effects is likely to be 
important, however, because the long-period Cepheids observed with \hst\
($P > 10$ days) arise from relatively massive ($M > 7~M_\odot$) and 
shortlived ($< 10^8$ yr) progenitors (CWC93), and variations 
in [O/Fe] are likely to be small in this young population (Wheeler, Sneden, 
\& Truran 1989).  There may be a larger absolute shift between the nebular 
and stellar abundance scales, but this is unimportant for the present 
application, as long as we calibrate and apply the abundance test in a 
self-consistent manner across the Key Project sample.

Figure 2 shows the HII region abundance distribution over the relevant 
range of radii in M101, with the locations of the Cepheid fields indicated 
by the horizontal lines.  The solid points are ``empirical" abundances 
based on the strengths of the reddening-corrected [OII]$\lambda$3727, 
[OIII]$\lambda$4959,5007, and H$\beta$ lines, taken from the survey of 
Kennicutt \& Garnett (1996), and calibrated following  ZKH.  The inner 
Cepheid field contains three HII regions measured in this survey, as 
indicated by the open circles.  The outer Cepheid field contains two 
bright HII regions that were not included in the Kennicutt \& Garnett 
(1996) survey.  Spectra for these objects have been obtained subsequently 
using the Blue Channel Spectrograph on the Multiple Mirror Telescope, and 
their abundances are also shown with open circles in Fig. 2.  The errors 
in the empirical abundances (shown in Fig. 2) are dominated by systematic 
uncertainty in the calibration of the absolute abundance scale, especially 
at high metallicities, where the calibration rests entirely on theoretical 
photoionization models.  Fitting to the empirical abundance distribution 
yields a difference between the inner and outer Cepheid fields of $+0.66 
\pm 0.20$ dex.

More robust abundances based on measured forbidden-line electron 
temperatures are available for a handful of HII regions in M101, and these 
data are shown as open triangles in Fig. 2.  The innermost point comes 
from Kinkel \& Rosa (1994), with the other data coming from Kennicutt \& 
Garnett (in preparation).  These data show a significant offset from the 
empirical abundances over part of the radial range, but the difference 
between the inner and outer fields is very similar, $+0.71 \pm 0.17$ dex, 
depending on whether the outermost points (NGC~5471) are included.  We 
adopt the average difference of  $+0.68 \pm 0.15$ dex.  Thus the inner and 
outer Cepheid fields span nearly a factor of five in abundance.

\subsection{\hst\ Photometry}

Observations of the M101 Cepheid fields were carried out in Cycles 3--5
for the outer field and Cycles 4--5 for the inner field.  Details of
the respective data sets, including complete discussions of the 
Cepheid photometry, variable identification, and period determinations
can be found in separate papers by Kelson \etal (1996) and Stetson \etal
(1997) for the outer and inner fields, respectively.  

Several technical obstacles prohibit a direct comparison of the inner and 
outer field photometry in their entirety. The bulk of the observations of 
the outer M101 field were obtained before the \hst\ refurbishment mission 
with the original Wide Field Camera (WFC), whereas the inner field was 
observed entirely in Cycles 4--5 with WFPC2.  Uncertainties in 
cross-calibrating WFC and WFPC2 could easily mask any real difference in 
the magnitudes and colors of the Cepheids (cf. Kelson \etal 1996).  
Fortunately we were able to obtain several exposures of the outer field 
with WFPC2, and these form the basis of the differential photometry 
analyzed here. To complicate matters further, most of the WFPC2 
observations were   obtained early in Cycle 4, when the operating 
temperature of the camera was adjusted to reduce charge transfer 
efficiency problems (Holtzman et al. 1995).  The WFPC2 data set includes a 
mix of ``warm" and ``cold" observations, and the zeropoint calibrations of 
these data may differ.  The combined uncertainties in tying together the 
WFC vs WFPC2 data and the warm vs cold WFPC2 data could easily amount to 
several hundredths of a magnitude in $V - I$ color, and mask (or mimic) 
any signature of a metallicity effect.

To avoid these problems we adopted a completely differential approach to 
the photometry of the inner and outer fields.  Our WFPC2 observations 
include 5 pairs of visits, when exposures of both fields were taken with 
the same filter, focus, and camera temperature.  These epochs (3 in $V$ 
and 2 in $I$) were used to define a common photometric scale for both 
fields.  Table 2 lists the journal of observations for this subset of the 
data.  The other (extensive) observations of each field were used to 
identify the Cepheid variables and measure their light curves, periods, 
and phases (Kelson \etal 1996; Stetson \etal 1997), but the magnitude 
scales used in this paper were determined solely from the 5 WFPC2 epochs 
listed in Table 2.

The primary set of photometry was performed with the DAOPHOT/ALLFRAME
package (Stetson 1994).  
Details of the reduction and calibration procedures
can be found in Hill \etal (1997) and Stetson \etal (1997). 
The inner and outer field data were reduced using identical point spread
functions, aperture corrections, and zeropoint calibrations applied
to the two data sets, to minimize the potential for any systematic
photometric offset.  As a check on our results we performed an 
independent reduction of the photometry using the DoPhot package
(Schecter, Mateo, \& Saha 1993).  To simplify the presentation  
we will first describe the main ALLFRAME analysis, followed by a
summary of the final results from both reductions.

\subsection{Cepheid Samples and Period-Luminosity Relations}

Cepheid variables were identified independently for the outer and inner 
fields by Kelson \etal (1996) and Stetson \etal (1997), respectively.  
Of the 29 Cepheids identified in the outer WFC field, 26 fell within the 
field of our WFPC2 images and were cleanly resolved in our ALLFRAME data. 
The periods of these stars range from $13 - 58$ days. For the inner field 
we identified a sample of 50 high-quality Cepheids with large amplitudes, 
well-determined light curves, and periods in the range $10 - 60$ days, and 
which appear to be free of crowding, as determined by the PSF fits and 
from visual inspection of the WFPC2 images (Stetson \etal 1997).  

The final product of the ALLFRAME run was 5 $V$ and 2 $I$ magnitudes for 
each star in the outer field, and 4 $V$ and 2 $I$ magnitudes for the  
inner field (some epochs contained cosmic ray split exposures).  These 
were averaged to produce intensity-weighted mean magnitudes 
for each Cepheid.  The average photometric errors returned by ALLFRAME for 
the stars in this magnitude range are $\pm 0.10$ mag in $V$ and $I$, and 
as low as $\pm 0.04$ for the brighter Cepheids.   
The largest source of random error in 
the magnitudes is the variability of the Cepheids themselves.  Because the 
data were taken at a small number of random phases, our average magnitudes 
will show a considerable scatter about the true mean values.  For the 
inner field data, we were able to reduce this problem by phase matching 
the observations to the full light curves and determining true mean 
magnitudes (but on the photometric zeropoint of the epochs in Table 2).  
This was not done for the outer field Cepheids, because of the long 
interval between the main WFC observations that define the light curves, 
and the larger photometric uncertainties in the WFC magnitudes.  We
confirmed that this procedure does not introduce any systematic
shift in the PL zeropoints.  

The resulting PL relations for the inner and outer fields are shown in 
Figures 3 and 4, respectively.  Superimposed in each case are the 
calibrating PL relations for the LMC, shifted to the best fitting distance 
modulus.  Least squares solutions, constrained to the slopes of the LMC PL 
relations, yield $V$ and $I$ moduli for the inner field of $\mu_V = 29.59 
\pm 0.06$ mag and $\mu_I = 29.43 \pm 0.05$ mag.  The corresponding 
solutions for the outer field are $\mu_V = 29.39 \pm 0.07$ mag and $\mu_I 
= 29.39 \pm 0.06$ mag.  Note that these solutions are based solely on
the subset of WFPC2 observations listed in Table 2, and are strictly
applicable only for the differential comparison of the inner and outer
fields presented here.  Fully 
calibrated magnitudes and PL fits for the fields can be found in 
Kelson \etal (1996) and Stetson \etal (1997). 

In the standard fitting procedure used in the Key Project, the difference 
in $V$ and $I$ moduli is assumed to represent the average reddening of the 
Cepheids; hence $E(V-I) = 0.16 \pm 0.03$ mag for the inner field and $0.00 
\pm 0.04$ mag for the outer field.  Applying the Cardelli \etal (1989) 
reddening curve yields for the true moduli $\mu_0 = 29.20 \pm 0.07$ mag 
for the inner field and $\mu_0 = 29.39 \pm 0.08$ mag for the outer field.  
These errors include only the uncertainty introduced by the dispersion of 
the PL relations (the errors in true modulus are comparable to those in 
$V$ and $I$ individually because the residuals in the two bandpasses are 
correlated). The errors in the absolute distance to M101 would be larger, 
because of additional uncertainties in the photometric zeropoints, the LMC 
distance and reddening, and other systematics, but these cancel out in the 
differential comparison.   Table 3 summarizes the results of these fits.  
The results quoted here refer to the complete Cepheid samples in each 
field, combined for the four WFPC2 detectors.  The fits for the complete 
Cepheid samples yield a difference in inferred true distance modulus of 
$-0.19 \pm 0.10$ mag, in the theoretically predicted sense (metal-rich 
closer).  

A possible source of concern in this comparison is the 
difference in the Cepheid period distributions between the two fields.  
Most of the Cepheids in the outer field have periods longer than 16 days 
(log $P \ge 1.2$), whereas the inner field Cepheids are more heavily 
weighted to the period range log $P = 1.0 - 1.4$.  The absence of 
short-period variables in the outer field is entirely an observational 
artifact, caused by interruptions in the observing sequence  combined with 
the brighter magnitude limit in the WFC data.  This difference in 
periods could introduce a spurious difference in distance moduli between 
the two  fields, if the slopes of the PL relations differ significantly 
from the canonical LMC values used in the fitting.  To minimize any bias 
from this effect we refitted the PL relations, restricting the data to 
Cepheids with log $P \ge 1.2$; this includes 24 of the 26 Cepheids in the 
outer field and 30 of the 50 Cepheids in the inner field.  In addition, we 
excluded 2 Cepheids in the inner field with very red colors ($V - I > 
1.5$).  Fits to these long-period subsamples yield true moduli for the 
inner and outer fields of 29.21 $\pm$ 0.09 mag and 29.34 $\pm$ 0.08 mag, 
respectively, or an inner $-$ outer difference of $-0.13 \pm 0.11$ mag 
(Table 3). The average distance modulus we derive for the outer field, 
29.36 $\pm$ 0.08 mag, is in excellent agreement with the value 29.34 $\pm$ 0.17 
mag (including all error terms) derived from the original photometry of 
the M101 Cepheids by Kelson \etal (1996). The outer field abundance is 
nearly the same as the LMC, which calibrates the PL relation, so its 
distance modulus is the appropriate one to apply to M101.

The same analysis was performed using the DoPhot photometry package
(Schechter, Mateo, \& Saha 1993).
A list of Cepheid candidates was generated independently,
and cross-checked with the final ALLFRAME list, yielding 34 Cepheids in
the inner field and 24 in the outer field.  The PL relations were then
fitted following the same procedures.  Fitting the complete Cepheid
sets ($P = 10 - 60$ days) yielded a inner $-$ outer difference in true distance
modulus of $-0.18 \pm 0.11$ mag ({\it vs} $-0.19 \pm 0.10$ mag for ALLFRAME).
A comparison using only Cepheids with $P = 16 - 60$ days yields a difference
of $-0.15 \pm 0.12$ mag ({\it vs} $-0.13 \pm 0.11$ mag for ALLFRAME).  
The two analyses
yield nearly identical values for the metallicity effect in M101.
At first glance it might appear that the agreement 
is {\it too} good, given the relatively large uncertainties,
but recall that the same stars are being measured, and the quoted uncertainties
largely reflect other factors, such as the intrinsic dispersion of the
PL relation, which affect both data sets uniformly.  However the consistency 
of the DoPhot vs ALLFRAME comparison does offer some assurance that 
the apparent difference in PL relations between the inner and outer
fields is not an artifact of field crowding errors in the data
reduction.

We adopt for our final result the average of these fits, for a net 
difference of $-0.165 \pm 0.10$ mag.  Combining this with the observed 
difference in oxygen abundance of $+$0.68 $\pm$ 0.15 dex (\S3.1) gives a 
metallicity dependence:
\begin{equation}
\gamma_{VI} = -0.24 \pm 0.16~{\rm mag~dex^{-1}} .
\end{equation}

\noindent
The quoted uncertainty includes errors in the Cepheid photometry and
in the abundance gradient.  Note that the measured $Z$-dependence is
only significant at the 1.5 $\sigma$ level.

\subsection{Interpretation}

Theoretical models predict that the Cepheid metallicity dependence 
is produced by a combination of brighter mean magnitudes and redder colors 
with increasing metallicity (e.g., CWC93).  Observational evidence for
systematic changes in the magnitudes and/or colors of Cepheids with
abundance has been offered by several authors, including
Caldwell \& Coulson (1986), Gieren \etal (1993), 
Laney \& Stobie (1994), Sasselov \etal (1997), and Kochanek (1997).  
With our $V$ and $I$ photometry alone, it is impossible to   
separate the effects of changes in intrinsic luminosity, intrinsic color, and
reddening.  However we can use the M101 results along those from other
Cepheid data sets to place useful limits on these contributions.

The nature of the shift in the PL relation between the M101 inner and
outer fields is illustrated in Figure 5, which shows the residuals of individual
Cepheids from the best fitting ALLFRAME $V$ and $I$ PL relations in each field.
The top and middle panels show the residuals for Cepheids in the outer and  
inner fields, respectively, relative to the respective PL fits in those 
fields.  The solid line in each case shows the expected trajectory due to 
variations in temperature (position in the instability strip), while
the dashed line shows the (nearly degenerate) trend expected from 
reddening variations.  The behavior of the residuals in M101 is 
similar to that seen in all of the Key Project data sets, insofar as 
most of the dispersion in the PL relations can be attributed to the 
finite strip width and to reddening variations.  The remaining scatter about
these correlations are presumably due to a combination of photometric
errors and uncertainties in deriving the mean magnitudes from a 
small number of epochs.

The bottom panel of Figure 5 shows the $V$ and $I$ residuals 
for stars in the inner field, but in this case with respect to the
mean PL relations in the {\it outer} field.  The outer field Cepheids 
have nearly the same abundances as the LMC Cepheids that calibrate the 
PL relation, so any metallicity dependence in the PL relation will 
appear as a systematic residual in this diagram.  The inner field
Cepheids actually show two distinct differences.  First there is a
general shift of the distribution to fainter magnitudes in both
$V$ and $I$, along the reddening/temperature trajectory.  This we 
tentatively attribute to a higher mean reddening in the inner field.
In addition there is a general shift of stars below the reddening
line, which implies that the mean colors of the inner field Cepheids
are redder (or else brighter) than would be expected from reddening
effects alone.  This represents the shift of $-0.16$ mag in
true distance modulus, transformed into the residual plane.  The 
clear appearance of this shift in Figure 5 confirms that
the difference in derived distance moduli between the inner and 
outer fields is significant, and is not
merely an artifact of scatter in the PL relations. 

We cannot distinguish from $V$ and $I$ data alone 
whether the shift in distance modulus is produced because the metal-rich 
Cepheids are brighter at both $V$ and $I$ (with no color change), 
because the metal-rich stars are intrinsically redder, or from some 
combination of luminosity and color changes.
However the M101 data place bounds on the variation in
either extreme.  The observed difference in true modulus of 0.16 mag
could be produced at constant color by a shift in both $V$ and $I$ moduli
of $-$0.16 mag (trivially), or by a shift in intrinsic color
of $+$0.08 mag.  Since the abundance difference in the two fields is 
$0.68 \pm 0.15$ dex, the implied metallicity dependences would be 
$-0.24 \pm 0.16$ \magdex\ in luminosity ($VI$), or $0.12 \pm 0.08$ 
\magdex\ in $V - I$ color.  In all likelihood some combination of
these effects is involved.  Note that the observed color excess in the
inner field, $E(V - I) = 0.16$ mag, is roughly twice as large
as any color difference that can be attributed to metallicity effects.
This means that part of the color difference between the Cepheids in
the inner and outer fields must be due to interstellar reddening.

Another way to constrain the effects of $Z$-induced color changes
on Cepheid distances is to test whether the observed Cepheid colors
correlate systematically with metal abundance.  Kochanek (1997) found
evidence for such a correlation, based on published photometry for
Cepheids in 17 galaxies.  We show an updated version of this test
in Figure 6.  Each point shows the average color excess for Cepheids  
in a given galaxy (or field), plotted as a function of metal abundance. 
The color excesses are derived from PL fits in $V$ and $I$ (equations [1]
and [2]), with the Galactic foreground reddening subtracted
(Burstein \& Heiles 1984).  The error bars signify the dispersion in 
the reddenings of individual Cepheids.  Data are taken from Kochanek (1997), 
along with data for IC~1613 (Freedman 1988), the two M101 fields from 
this paper (open circles), and the three fields in M31 (Freedman \& Madore 
1990), shown as crosses.  We excluded galaxies in 
the Kochanek sample with color excesses derived at wavelengths other 
than $V$ and $I$, and galaxies without measured HII region abundances. 
This leaves a sample
of 19 fields in 16 galaxies.  

Figure 6 shows a clear trend between Cepheid color excess and 
metallicity, confirming Kochanek (1997).  The best fitting slope
to our data is $\delta E(V - I)/\delta[O/H] = 0.12 \pm 0.08$ \magdex\ 
(a linear fit in $Z$ yields virtually
the same slope over the relevant abundance range).  This color gradient
is similar to what is inferred above from the M101 inner and outer fields,
but the trend in Figure 6 includes any changes in mean interstellar 
reddening with metallicity.  It is interesting that the color trend is
restricted to galaxies with
abundances higher than that of the LMC ([O/H] = $-0.4$), where one
might expect a higher dust-to-gas ratio and thus a higher reddening.
The absence of a color trend at lower abundances suggests that interstellar
reddening may well be responsible for much if not most of the observed
color trend.  Even if all of the trend in Fig. 6 were due to metallicity
effects, it would introduce at most a bias in the distance modulus
$\gamma_{VI} \simeq -0.25 \pm 0.17$ \magdex.

\section{Comparison with Other Studies}

The first application of the differential metallicity test for 
Cepheids over a range of galactocentric distances was carried out 
for M31 by Freedman \& Madore (1990; hereafter FM90).  FM90  
analyzed random-phase $BVRI$ observations for 38 Cepheids in 
three fields in M31, located at radii of 3, 10, and 20 kpc.  
No difference in true distance modulus was detected for the 3 and 10 kpc 
fields, but the derived true modulus for the 20 kpc field was higher
by $0.25 \pm 0.17$ mag, when determined from the $V$ and $I$ data alone.
Combining this with an estimated abundance gradient of 0.75 dex over
3--20 kpc yielded a metallicity dependence $\gamma_{BVRI} = -0.32 \pm 0.21$ 
\magdex\ and $\gamma_{VI} = -0.39 \pm 0.26$ \magdex.   FM90 did not    
solve for $\gamma$ explicitly, and the numbers listed here were 
derived from the PL solutions and abundance gradients given in their paper,
following Gould (1994).  
In view of the low statistical significance of this result, FM90 
interpreted it as a nondetection, and consistent with no measureable 
$Z$ dependence of the PL zeropoint.

The M31 data were subsequently reanalyzed by Gould (1994), who
used the correlations between magnitude residuals in $BVRI$ in an
attempt to separate the effects of metallicity and reddening.  His
best solutions for the same data yielded values $\gamma_{BVRI} =  
-0.88 \pm 0.16$ \magdex\ and $\gamma_{BVI} = -0.56 \pm 0.20$ \magdex,
a much larger and statistically significant dependence.  These results
cannot are not directly comparable to ours, because the wavelength baseline
extends to the blue, where the metallicity effects are expected to be 
larger.  However the analysis  
illustrates the large range of dependences which are consistent with the 
M31 data.

Unfortunately both of these analyses were based on an erroneous
value for the abundance gradient in M31.  FM90 quoted an abundance range
of 0.75 dex over the three Cepheid fields (1.7 to 0.3 solar), citing 
Blair, Kirshner, \& Chevalier (1982), but the abundance gradient given
in the latter paper corresponds to a range
of only 0.44 dex, when placed at the FM90 distance.  Adding data from
Dennefeld \& Kunth (1981) and recalibrating the oxygen abundances on the
ZKH scale lowers the abundance range further to 0.31 $\pm$ 0.16 dex
over the 3 Cepheid fields (from 1.8 to 0.9 solar).  Applying this 
correction to the FM90 PL solutions 
changes their metallicity dependence to $\gamma_{VI} = -0.94 \pm 0.78$
\magdex, with the larger fractional error reflecting the additional 
uncertainty in the abundance gradient.  Likewise the revised 
Gould (1994) solutions become   $\gamma_{BVRI} = -2.1 \pm 1.1$
\magdex\ and $\gamma_{BVI} = -1.4 \pm 0.8$ \magdex.  
We conclude, in agreement with FM90, that the existing
M31 data do not offer very stringent constraints on the Cepheid 
metallicity dependence.  Ongoing efforts to obtain more precise photometry over
a wider wavelength baseline may improve the constraints  
(Kaluzny \etal 1997; Freedman \etal 1997).

A different approach to constraining the metallicity dependence
has been employed by Beaulieu \etal (1997), Sasselov \etal (1997) 
and Kochanek (1997).  Beaulieu \etal and 
Sasselov \etal have compiled a large database of photometry of
LMC and SMC Cepheids from the EROS microlensing project, and 
applied a differential multivariate analysis to determine the
maximum likelihood change in distance moduli and color with 
metallicity.  They derive $\gamma_{VI} = -0.44{^{+0.1}_{-0.2}}$ \magdex,
This was derived by transforming photometry made at slightly shorter 
wavelengths (corresponding approximately to Johnson $V$ and $R$), 
so the dependence might be slightly steeper than one would
derive from $V$ and $I$ photometry. 

Kochanek (1997) has applied a similar approach to Cepheid data for 
17 galaxies measured from the ground and with \hst, solving 
independently for residual extinction and metallicity-dependent
colors and luminosities.  His maximum likelihood 
solution yields a mean change in magnitude at $V$ and $I$ of 
$-0.14 \pm 0.14$ \magdex\ and a $V - I$ color change of 0.13 $\pm$ 0.04
\magdex.  Taken together these would introduce a metallicity 
dependence in inferred true modulus of $\gamma_{VI} \sim -0.4 \pm 0.2$ \magdex.
These dependences are larger than the $\gamma_{VI} = -0.24 \pm 0.16$ \magdex\
derived here from our M101 observations, but the results are 
consistent within their quoted uncertainties.

\section{External Tests}

The metallicity sensitivity of the Cepheid distance scale can
be tested externally if independent distances are available  
for galaxies spanning a large range in abundance.  The tip of the
red giant branch (TRGB) method offers a strong test of the Cepheid
scale over a large abundance
baseline.  Two other secondary distance indicators,
the planetary nebula luminosity function (PNLF) and the Tully-Fisher
(TF) method, provide indirect constraints on the magnitude of the
$Z$ effects as well.  

\subsection{Comparison with TRGB Distances}

The TRGB method utilizes the nearly constant luminosity of the red
giant branch tip in the $I$ band ($M_I = -4.0 \pm 0.1$).  Globular 
cluster observations by Da Costa \& Armandroff (1990) 
indicate that the tip luminosity remains constant
over a large range of metallicity ($-2.2 < [Fe/H] < -0.7$), 
and theoretical isochrones show a similar constancy over the same
abundance range, and for ages of 
2 $-$ 15 Gyr (Lee, Freedman, \& Madore 1993; Madore, Freedman, 
\& Sakai 1997, and references therein).  The method has been applied 
to 9 galaxies with independently determined Cepheid distances,
as summarized in Table 4.  Included are  
Cepheid and TRGB distances with their associated uncertainties and the 
metal abundances of the galaxies.  We list two sets of Cepheid distances,
the published values and those determined from the PL relations at 
$V$ and $I$ alone, if different.  Testing for a $Z$ dependence in the
$VI$ Cepheid data alone is much cleaner, as it avoids problems introduced
by a wavelength-dependent metallicity dependence, though the distance
moduli determined from $V$ and $I$ alone are often less precise.  
The abundances listed in Table 4 are   
determined from HII regions (Skillman, Kennicutt, \& Hodge 1989; ZKH), 
and apply to the Cepheids (not the red giants),
because we are testing for metallicity effects in the Population I Cepheids.
The galaxies span 7 magnitudes in absolute magnitudes and 
a factor of 50 in (Pop I) metal abundance, thus providing us with a sensitive
test for metallicity effects in the Cepheids.  The red giant abundances
of the galaxies all lie within the calibrated range of the TRGB method,
where metallicity effects in the TRGB distances should be small.

We also include an indirect comparison for two spirals in the
Leo I group, NGC~3351 and NGC~3368.  These galaxies have \hst\ Cepheid 
distances measured from Graham \etal (1997) and Hjorth \& Tanvir (1997),
respectively, and a TRGB distance determined for an elliptical member of the
group, NGC~3379, from Sakai \etal (1997b).  In these cases
the comparison may be influenced by group depth, so the data should be
accorded lower weight.  The TRGB distances for these objects are marked
in parentheses in Table 4.

The absolute distances measured with the two methods
show good agreement, with an rms scatter of $\pm$0.16 mag (8\% in distance).
Figure 7 shows the residuals plotted as a function of metal abundance
(NGC~3351 and NGC~3368 are indicated with open circles).
There is a slight trend with abundance, and a least squares
fit yields a best fitting slope $\gamma_{VI} = -0.16 \pm 0.08$ \magdex, shown
by the solid line in Figure 7.  Comparing to the more heterogeneous set
of published Cepheid distances (not all at $V$ and $I$ alone) 
yields a nearly identical fit ($\gamma =
-0.14 \pm 0.08$).  

One might worry that these trends could be influenced
by a metallicity dependence in the TRGB distances (cf. Saleris \& Cassissi
1997).  This is very unlikely, however, because the red giant abundances in
these galaxies are virtually uncorrelated with the metallicites of the
Cepheid fields, and changing them does not appreciably influence the 
slope of the relation in Figure 7.  
As a check we recalculated the TRGB distances using a
steeper $Z$ dependence for the RR Lyrae luminosity scale, 
$dM_V(RR)/d[Fe/H] = 0.30$ instead of 0.17 as adopted in Table 4.  Repeating
the comparison in Figure 7 with the new TRGB distances yielded nearly
the same result as before, $\gamma_{VI} = -0.12 \pm 0.08$ \magdex, 
and if anything produces 
a slightly weaker Cepheid $Z$ dependence.  Lee \etal (1993) directly
compared TRGB and Cepheid distances as a function of giant branch abundance,
and found no significant dependence on (Pop II) metallicity.

Although our analysis has focussed on the $Z$ dependence of the PL relation in
$V$ and $I$, seven of the galaxies in Table 4 have Cepheid distances
based on $BVRI$ fits, and we can check whether there is evidence for
a significant metallicity dependence in those distances.  Repeating
the comparision yields very similar results, with 
$\gamma_{BVRI} = -0.13 \pm 0.11$.  Given the small sample and the
inhomogeneity of the data set, we regard this as a tentative result.
However it is interesting that there is no indication in the TRGB
comparison for a strong metallicity dependence, even in the blue.

To summarize, the comparison of Cepheid and TRGB distances yields another
marginal detection of a Cepheid metallicity dependence, $\gamma_{VI} 
= -0.16 \pm 0.08$ \magdex, which is 
formally consistent with our M101 results ($\gamma_{VI} = -0.24$),
as well as with the theoretically predicted dependence 
($\gamma _{VI}= -0.11$).  The dashed and dotted lines in Figure 7 show 
the dependences that are expected for values of $\gamma$ between
$-0.24$ and $-0.88$; the strong dependences suggested by 
previous studies are not confirmed here.   These results, 
together with our M101 analysis, offer the strongest 
evidence for a weak metallicity dependence to the PL relations in 
$V$ and $I$.

\subsection{Other Tests}

We can perform a similar check for galaxies with Cepheid and PNLF
distances.  Direct comparisons are available
for 7 galaxies, and indirect comparisons via galaxies in the same
group or cluster are available for 4 other Cepheid hosts.  The data
are listed in Table 4, with the indirect comparisons in parentheses.
Most of these data were taken directly from 
comparisons in Soffner \etal (1996) and Feldmeier, Ciardullo, \& Jacoby (1997),
with the addition here of abundance values appropriate to the 
Cepheid fields.

Figure 8 shows the Cepheid $-$ PNLF distance modulus residuals as a
function of metal abundance.  Again there is an excellent correlation 
between the absolute distances ($\pm$0.19 mag rms), 
as shown previously by Soffner \etal (1996) and Feldmeier \etal (1997).
This comparison is not as clean as the TRGB test, 
because the galaxies cover a smaller range in
metal abundance, and the PNLF method itself
may have a metallicity dependence at the $\pm$0.2 mag level 
(Ciardullo \& Jacoby 1992).  Given these large uncertainties, the PNLF
distances do not impose tight constraints on the Cepheid metallicity
dependence, and if anything Figure 8 is suggestive of a modest
$Z$ dependence in the PNLF distances.  

As a final check we can use the slope of the local Tully-Fisher (TF)
relation to constrain the magnitude of any Cepheid $Z$ dependence.
Distances from the TF method are currently available 
for 15 Cepheid calibrating galaxies.  Unfortunately  
a residual test of the sort applied to the TRGB and PNLF distances
cannot be applied here, because 
the intrinsic dispersion of the TF relation is much larger, and
because one of the parameters
in the TF method, galaxy luminosity, is itself strongly correlated
with disk metallicity (Garnett \& Shields 1987; ZKH).  
As a result any metallicity dependence in the Cepheids will primarily
act to change the apparent slope of the TF relation, rather than affect the
scatter of the relation (Gould 1994).  We can exploit this effect,
however, and test whether the slope of a TF relation calibrated
from Cepheid distances is significantly shallower than the slope
derived from observations of distant spirals, where relative distances
are derived independently of the Cepheid scale.  The magnitude of
the slope change for a given Cepheid $Z$-dependence can be easily
derived.  Consider for example the $I$-band TF relation:
\begin{equation}
M_I = {a~{\rm log} W + M_I(0)}
\end{equation}

\noindent
If we define the metallicity dependence $\gamma$ in the usual way,
\begin{equation}
\gamma = {{d(m - M)} \over {d[O/H]}}
\end{equation}

\noindent
and define $\beta$ as the slope of the absolute magnitude vs 
metallicity relation,
\begin{equation}
\beta = {dM_I \over d[O/H]}
\end{equation}

\noindent
then the effect of the Cepheid $Z$ dependence will be to
alter the TF slope:
\begin{equation}
{{\delta a} \over a} = {- {\gamma \over \beta}}
\end{equation}

We used the $I$-band TF analysis of Giovanelli \etal (1997) to derive
representative values of $a$ and $\beta$.  These authors   
compiled observations of 555 spiral galaxies in 24 rich clusters and
applied incompleteness corrections to 
define the slope of the TF relation independent of the Cepheid scale,
yielding $a = -7.67 \pm 0.11$ (bivariate fit).  
Giovanelli \etal also compiled
$I$ magnitudes and linewidths for 15 galaxies with Cepheid distances,
including 12 with measured HII region abundances. 
A least squares fit of absolute magnitude vs
[O/H] yields a luminosity-metallicity slope $\beta$ = 4.8 $\pm$ 1.0.
This is similar to $\beta$ = 4.6 from ZKH (but using $B$ magnitudes).
Applying $a = -7.67$ and $\beta = 4.8$ to equation (8) then allows
us to predict the TF slopes for the Cepheid calibrating galaxies,
for various values of $\gamma$.  The predicted changes in slope
range from 5\% (to $a = -7.3$) for $\gamma = -0.24$ \magdex\ 
to 18\% ($a = -6.3$) for $\gamma = -0.88$ \magdex.

Figure 9 shows the TF relation for the 15 Cepheid 
calibrating galaxies (Giovanelli \etal 1997), with the predicted
slopes superimposed.  The slope of the Cepheid-derived TF relation is 
actually {\it steeper} than the cluster-calibrated relation.
A bivariate fit to the 12 galaxies with reliable linewidths 
(Giovanelli \etal 1997) yields $a = -8.65 \pm 0.66$.  Fitting 
all 15 galaxies yields an even steeper slope.    
The slopes of the Cepheid-calibrated relations thus are consistent with
no $Z$ dependence, though the large uncertainty in the slope, not unexpected 
for a small sample, again limits the usefulness of this test.  
Our slope also agrees with that derived by Pierce \& Tully (1988)
for the Ursa Major cluster ($a_I = -8.72$), but their linewidths
and magnitudes were calibrated using slightly different methods, so we
cannot compare our slope directly to theirs.  
Nevertheless the Tully-Fisher data offer further evidence against a 
large abundance
effect in the Cepheids ($\gamma_{VI} < -0.5$).  It should be possible 
to tighten these limits as more \hst\ Cepheid distances are accumulated
over the course of the Key Project.

\section{Discussion}

We begin by summarizing the results of the various tests presented
in this paper.  The direct comparison of PL relations for the inner and outer
Cepheid fields in M101 imply a difference in inferred true
distance moduli of 0.16 $\pm$ 0.10 mag (inner field closer),
over an abundance baseline of 0.68 $\pm$ 0.15 dex.  This 
implies a metallicity dependence of distance modulus inferred
from $V$ and $I$ PL relations of $\gamma_{VI} = -0.24 \pm 0.16$ \magdex.
An external comparison of Cepheid distances with those obtained
using the TRGB  method implies an even smaller
dependence:  $\gamma_{VI} = -0.16 \pm 0.08$ \magdex, over a metallicity
range of 1.7 dex.  Comparisons with two sets of distance estimates,
using the PNLF and Tully-Fisher methods, are consistent with no
systematic metallicity dependence within large uncertainties, and
they limit the magnitude of any dependence to $\gamma_{VI} \gg -0.5$ \magdex.
A comparable upper limit is implied by the systematic dependence
of Cepheid color on metal abundance.   
All of these results are consistent with a 
maximum range in $\gamma_{VI} \simeq -0.25 \pm 0.25$ \magdex, when
measured from PL relations in $V$ and $I$.
This range is consistent with the very weak sensitivity predicted by
theory ($\gamma_{VI} \simeq -0.1$ \magdex), or with no dependence at all.

It is important to emphasize that the
results quoted here apply only to PL relations measured in the 
$V$ and $I$ bands.
There is theoretical and observational justification for suspecting that 
the metallicity dependence at bluer wavelengths is considerably stronger,
but this is not relevant to the current \hst\ measurements, which
are carried out exclusively at $V$ and $I$.  Until the magnitude of the
$Z$ dependence at bluer wavelengths is well calibrated, it would be
prudent to restrict extragalactic applications of Cepheids to 
observations at $V$ and longer wavelengths.

What are the implications for the distance scale and \h0\ of a 
metallicity dependence in this range?  The abundance effects
clearly are most important for 
galaxies with unusually high or low abundances relative
to the LMC, which serves as the zeropoint calibrator for the 
PL relation.  On the ZKH calibration the LMC has an oxygen abundance
12 $+$ log(O/H) = 8.50, or [O/H] = $-0.40$.  The most metal-rich 
galaxies in the Key Project sample reach [O/H] $\simeq$ 0.3 
(e.g., NGC~3351, M100, M101 inner field).  For $\gamma = -0.24$ \magdex\ 
the corresponding error in distance is $-$9\% ($-$0.18 mag), in
the sense that distances to the metal-rich galaxies are systematically
underestimated.  For $\gamma = -0.5$ \magdex\ 
the distance change is as much as $-$18\%, but this is a hard upper limit,
applied to the most metal-rich galaxies observed with \hst, and 
computed for the maximum allowable metallicity dependence.  It is 
worth noting that any strong Cepheid metallicity dependence would tend
to cause the Cepheid distance to the LMC to be {\it overestimated},
and compensate partly for the distance effects on more distant metal-rich
galaxies.

The Cepheid $Z$ dependence tends to cause the distances of metal-poor galaxies
to be overestimated.  The abundances of the most metal-poor galaxies in the Key
Project are only slightly below that of the LMC 
(e.g., NGC~3319, with [O/H] $\sim -0.5$).  For those galaxies the 
predicted metallicity effects are less than 3\%, even for the largest 
reasonable values of $\gamma$.  The effects may be larger for some of 
the SN Ia calibrators being observed by Sandage and collaborators
with \hst\ (e.g., Saha \etal 1994).  Beaulieu \etal (1997) and 
Kochanek (1997) have suggested that the value of \h0\ derived from the 
SN Ia method may increase by 16 $-$ 24\%, for an assumed 
$\gamma_{VI} \simeq -0.4$ 
\magdex.  Our analysis suggests a much smaller difference, based on the 
weaker metallicity dependence observed in M101, and the likelihood that the
abundances of the SN~Ia calibrators are closer to the
LMC than has been assumed by the other authors.  The most metal-poor SN~Ia
calibrator is probably NGC~5253, with [O/H] $\simeq -0.7$ (Webster
\& Smith 1983), and its distance would have been overestimated by only 
0 $-$ 8\%, for the range of $\gamma$ considered here.  

Given the probable magnitude of these effects, should one apply 
metallicity corrections to Cepheid distances measured
with \hst?  As illustrated above, the effects on individual distances
are probably 10\% {\it at most}, and for most galaxies 
the changes are at the few percent level.
Cepheid distances currently measured with \hst\ are subject to
several random and systematic uncertainties at the 5\% level, including
uncertainty in the LMC distance, the absolute calibration of the
$V$ and $I$ magnitudes measured with WFPC2, and uncertainties in the
reddening corrections applied to the WFPC2 photometry (cf. Ferrarese
\etal 1996; Hill \etal 1997; Madore \etal 1997).  In view of these
other errors and the uncertain magnitude of the metal abundance correction
itself, we believe that it is most prudent at this time 
not to apply a metallicity
correction to individual Cepheid distances, but rather to include
this source of uncertainty in the systematic error budget for the distance.
At the end of the Key Project we hope to have better estimates of these 
uncertainties, and a much improved zeropoint calibration of the PL relation
(Kennicutt \etal 1995).  

In the meantime, one can assess the importance of metal abundance effects
on the overall distance scale.  Beaulieu \etal (1997), Sasselov \etal (1997) 
and Kochanek (1997) explored the consequences of a $\gamma = -0.44$ \magdex\ 
dependence on currently published measurements of \h0, and
concluded that the Key Project \h0\ value could be 
reduced by up to 10\%, based on the Virgo cluster calibration of
Freedman \etal (1994b) and Mould \etal (1995).  However the most
recent Key Project \h0\ determination is based on a much larger 
sample of Cepheid distances and secondary distance indicators, and 
is much less susceptible to this metallicity bias (Madore \etal 1997).
As an illustration,  
Figure 10 shows a histogram of oxygen abundances for the Key Project
galaxies with HII region measurements.  These include the \hst\
target galaxies, with abundances from ZKH, and the groundbased
calibrators M31, M33, NGC~300, NGC~2366, NGC~2403, and NGC~3109.
Although there is a spread of over an order of magnitude in these
abundances, the median value is [O/H] $\simeq -0.3$, 
nearly identical to the LMC abundance of [O/H] = $-0.4$.  Thus the
effect of even a substantial Cepheid metallicity dependence will be
negligible for any distance indicator that is calibrated using most or all
of the Key Project Cepheid sample (e.g., the Tully-Fisher relation).
The net abundance effect will differ for each secondary
distance method, depending on the subset of Cepheid calibrating
galaxies, but for most methods the mean Cepheid abundance will be solar
or below, so the $Z$ effects should be at the
0.10 mag level or less (less than 5\% in distance and \h0).  The situation  
is less certain for the SN~Ia distance scale,  
because abundances for most of the Cepheid calibrators 
have not yet been measured.  We are obtaining HII region spectra
for the galaxies in order to assess the importance of abundance 
effects on the SN~Ia distance scale.

Systematic metal abundance effects may be more important 
for galaxies in the Virgo and Fornax clusters.
It is well established that spirals in the core of the Virgo
cluster are systematically more metal-rich than their field
counterparts (Skillman \etal 1996), and this region contains 
several Cepheid calibrating galaxies (M100, NGC~4548, NGC~4535,
NGC~4639, NGC~4571).  Less is known about the abundances in 
the Fornax cluster, but the Key Project sample includes 
three Fornax members (NGC~1365, NGC~1425, NGC~1326A), and the
one spiral with measured abundances (NGC~1365) is relatively metal-rich.
It will be important when calibrating \h0\ via galaxies in
Virgo and Fornax to consider the possible systematic abundance
effects.  However from the comparisons given earlier the maximum
magnitude of an abundance effect will be to underestimate the
distances by $\sim$10\%, and consequently lead to a decrease in 
\h0\ of 10\%.

Our $V$ and $I$ photometry of the two M101 fields provides tentative 
evidence for a small metallicity dependence of the PL relation at these
wavelengths, and when combined with other tests, places firm
upper limits on the magnitude of the metallicity effect.  These are
sufficient to ensure that the overall consequences of a metallicity
dependence on the ultimate \h0\ calibrations with \hst\ are small (at the
level of a several percent 
or less), and comparable to a multitude of other known error sources in the
distance scale calibration.  However the tests presented in this 
paper fall far short of establishing a definitive calibration of the
PL metallicity dependence.  Such a calibration would
significantly improve the accuracy of individual Cepheid distance
determinations to individual galaxies, and provide valuable constraints
on the theoretical understanding of the Cepheids themselves.  Here we briefly
describe several observations that would address this need.  We also
refer to reader to Kochanek (1997) for a discussion of this subject.

As has been discussed previously by Madore \& Freedman (1985) and
Kochanek (1997), the primary limitation of the \hst\ observations 
is the $V - I$ wavelength baseline.  This restriction was imposed primarily
by the need to maximize observing efficiency on \hst, and by the 
absence, until recently, of a near-infrared imaging capability on \hst.
A targeted program aimed at obtaining high precision Cepheid photometry
of a single galaxy such as M101 over a wide wavelength baseline and 
over a wide range of abundances would enable one to 
break the degeneracy between reddening and metallicity
effects, and accurately calibrate the $Z$ dependence.  Groundbased
observations of this kind are being obtained by at least two groups
for M31.  The installation of 
NICMOS on \hst\ makes such a program very feasible in a more distant
galaxy with a steeper abundance gradient, such as M101.

The effects of metallicity and reddening on the PL relation are
smaller at near-infrared
wavelengths (e.g., McGonegal \etal 1982; CWC93),
which argues strongly for supplementing 
the existing WFPC2 $V$ and $I$ photometry of Cepheid calibrating
galaxies with NICMOS imaging in $H$
and/or $J$.   
As the absolute 
uncertainty in \h0\ approaches the 10\% level, the systematic
uncertainties associated with metallicity and reddening corrections
to the Cepheid distances will become the dominant source of error in the
entire extragalactic distance ladder.  Thus a modest effort to 
assess and correct these systematic errors will have great leverage
in improving the calibration of individual extragalactic distances,
and ultimately the Hubble constant as well.

\acknowledgments
We are pleased to thank George Jacoby, Chris Kochanek, and
Dimitar Sasselov for useful comments and discussions.  We also
thank the referee for suggesting several improvements to the paper.
The work presented in this paper is based on observations made by the
NASA/ESA Hubble Space Telescope, obtained by the Space Telescope
Science Institute, which is operated by AURA, Inc. under NASA contract
No. 5-26555.  We gratefully acknowledge the support of the NASA and
STScI support staff, with special thanks to Peggy Stanley and Doug
Van Orsow.  Support for this work was provided by NASA through grant
GO-2227-87A from STScI.  This 
research has made use of the NASA/IPAC Extragalactic Database (NED)
which is operated by the Jet Propulsion Laboratory, California
Institute of Technology, under a contract with the National
Aeronautics and Space Administration.

\newpage

\centerline{{\bf Figure Captions}}

FIG. 1.--- POSS image of M101 from the Digital Sky Survey, with the
locations of the two WFPC2 fields indicated.  North is up, east is to
the left.

FIG. 2.--- Radial abundance distribution in M101, as derived from HII
region spectra.  The round points and circles are empirical abundances
based on the calibration of ZKH.  Open circles denote HII regions that
are located in the inner and outer Cepheid fields.  Open triangles 
denote abundances based on HII regions with measured electron temperatures.
The horizontal lines indicate the radial coverage of the Cepheid fields.

FIG. 3.--- Observed Cepheid PL relations from ALLFRAME photometry of 
the M101 outer field in 
$V$ (top) and $I$ (bottom).  The lines show the least squares fits,
constrained to the slope of the LMC calibrating relations.

FIG. 4.--- Observed Cepheid PL relations in the M101 inner field.
Notation the same as in Fig. 3.

FIG. 5.--- Distributions of $V$ and $I$ residuals of individual Cepheids 
from the PL fits shown in Figs. 2 and 3.  The top panel shows residuals
of the outer field Cepheids from the outer field fit.  The middle 
panel shows the residuals of inner field Cepheids from the inner field
fit.  The bottom panel shows the residuals of the inner field Cepheids
from the outer field PL fits.  In each panel the solid line shows the
expected residuals from instability strip width effects, while the 
dashed line shows the expected trend from reddening variations.

FIG. 6.--- Correlation of the average color excess of the Cepheids
in a given galaxy or field with abundance.  In each case the Galactic
foreground reddening has been subtracted.  The solid line show the
bivariate least squares fit to the correlation.

FIG. 7.--- Difference between the distance moduli of nearby galaxies
determined from Cepheids and by the TRGB method, plotted as a function
of the Cepheid abundances.  Solid points indicate
direct comparisons of galaxies with Cepheid and TRGB distances, while
the open circles indicate indirect comparisons using Cepheid and TRGB
distances of different galaxies in Leo I group.  The solid line shows
a bivariate least squares fit to the relation.  The dashed and dotted lines show
trends expected for a different Cepheid $Z$ dependences. 

FIG. 8.--- Similar comparison to Fig. 7, but in this case comparing
Cepheid distances with those derived from the PNLF method.  Solid
points indicate direct comparisons in galaxies with Cepheid and
PNLF distances, while open circles are indirect comparisons using
different galaxies in the same group.  The lines show expected trends
as in Fig. 6.

FIG. 9.--- The $I$-band Tully-Fisher relation for Cepheid calibrating
galaxies, using data from Giovanelli \etal (1997).  The open circles
indicate galaxies with poorly determined linewidths, and which are 
not included in the fit.  The solid line shows the best fitting relation.
The dashed and dotted lines show the relations expected for different
Cepheid $Z$ dependences $\gamma$, as in Figs. 6 and 7.

FIG. 10.--- Distribution of Cepheid abundances in the Key Project 
sample, as determined from HII region measurements, on the calibration
of ZKH.  The vertical line indicates the abundance of the LMC, which
defines the PL calibrations.


\newpage
\begin{references}
\reference{}{Beaulieu, J.P. 1997, \aap, 318, L47}
\reference{}{Blair, W.P., Kirshner, R.P., \& Chevalier, R.A. 1981, \apj, 247, 879}
\reference{}{Burstein, D., \& Heiles, C. 1984, \apjs, 54, 33}
\reference{}{Caldwell, J.A.R., \& Coulson, I.M. 1986, \mnras, 218, 233}
\reference{}{Capaccioli, M., Piotto, G., \& Bresolin, F. 1992, \aj, 103, 1151}
\reference{}{Cardelli, J.A., Clayton, G.C. \& Mathis, J.S. 1989, 
\apj, 345, 245}
\reference{}{Ciardullo, R., \& Jacoby, G.H. 1992, \apj, 388, 268} 

\reference{}{Chiosi, C., Wood, P.R., \& Capitanio, N. 1993,
\apjs, 86, 541 (CWC93)}
\reference{}{Chiosi, C., Wood, P.R., \& Capitanio, N. 1997, in preparation (CWC97)}
\reference{}{Da Costa, G.S., \& Armandroff, T.E. 1990, \aj, 100, 162}
\reference{}{Dennefeld, M., \& Kunth, D. 1981, \aj, 86, 989}
\reference{}{Feldmeier, J.J., Ciardullo, R., \& Jacoby, G.H. 1997, \apj, 479, 231}
\reference{}{Ferrarese, L. \etal 1996, \apj, 464, 568}
\reference{}{Freedman, W.L. 1988, \apj, 326, 691}
\reference{}{Freedman, W.L., \& Madore, B.F. 1990, \apj, 365, 186 (FM90)}
\reference{}{Freedman, W.L., Wilson, C.D., \& Madore, B.F. 1991, \apj, 372, 455}
\reference{}{Freedman, W.L. \etal 1992, \apj, 396, 80}
\reference{}{Freedman, W.L. \etal 1994a, \apj, 427, 628}
\reference{}{Freedman, W.L. \etal 1994b, Nature, 371, 757}
\reference{}{Freedman, W.L., Madore, B.F., \& Sakai, S. 1997, in preparation}
\reference{}{Garnett, D.R., \& Shields, G.S. 1987, \apj, 317, 82}
\reference{}{Gieren, W.P. 1993, \mnras, 265, 184}
\reference{}{Giovanelli, R., Haynes, M.P., da Costa, L, Freudling, W.,
Salzer, J., \& Wegner, G. 1997, \aj, 113, 53}
\reference{}{Graham, J.A. \etal 1997, \apj, 477, 535}
\reference{}{Gould, A. 1994, \apj, 426, 542}
\reference{}{Hill, R. \etal 1997, \apj, in press}
\reference{}{Hjorth, J., \& Tanvir, N.R. 1997, \apj, 482, 68}
\reference{}{Holtzman, J. \etal 1995, \pasp, 107, 1065}
\reference{}{Kaluzny, J., Stanek, K.Z., Krockenberger, Z.Z., Sasselov, D.D., Tonry, J.L., \& Mateo, M. 1997, \aj, in press}
\reference{}{Kelson, D. \etal 1996, \apj, 463, 26}
\reference{}{Kennicutt, R.C., Freedman, W.L., \& Mould, J.R. 1995, \aj, 110,1476}
\reference{}{Kennicutt, R.C., \& Garnett, D.R. 1996, \apj, 456, 504}
\reference{}{Kennicutt, R.C., \& Bresolin, F. 1997, in preparation}
\reference{}{Kinkel, U., \& Rosa, M.R. 1994, \aap, 282, L37}
\reference{}{Kochanek, C.S. 1997, \apj, in press}
\reference{}{Laney, C.D., \& Stobie, R.S. 1994, \mnras, 266, 441}
\reference{}{Lee, M.G., Freedman, W.L., \& Madore, B.F. 1993, \apj, 417, 553}
\reference{}{Madore, B.F., \& Freedman, W.L. 1985, \aj, 90. 1104} 
\reference{}{Madore, B.F., \& Freedman, W.L. 1991, \pasp, 103, 933}
\reference{}{Madore, B.F., Freedman, W.L., \& Sakai, S. 1997, in The 
Extragalactic Distance Scale, ed. M. Livio, M. Donahue, \& N. Panagia 
(Cambridge: Cambridge Univ. Press), 239}
\reference{}{Madore, B.F. \etal 1997, Nature, submitted}
\reference{}{McGonegal, R., McLaran, R.A., McAlary, C.W., \& Madore, B.F. 
1982, \apj, 257, L33}
\reference{}{McMillan, R., Ciardullo, R., \& Jacoby, G.H. 1993, \apj, 416, 62}
\reference{}{Mould, J. \etal 1991, \apj, 383, 467}
\reference{}{Oey, M.S., \& Kennicutt, R.C. 1993, \apj, 411, 137}
\reference{}{Pierce, M.J., \& Tully, R.B. 1988, \apj, 330, 579}
\reference{}{Richer, M.G., \& McCall, M.L. 1995, \apj, 445, 642}
\reference{}{Saha, A. \etal 1994, \apj, 425, 14}
\reference{}{Saha, A. \etal 1995, \apj, 438, 8}
\reference{}{Sakai, S., Madore, B.F., \& Freedman, W.L. 1996, \apj, 461, 713}
\reference{}{Sakai, S., Madore, B.F., \& Freedman, W.L. 1997a, \apj, 480, 589}
\reference{}{Sakai, S., Madore, B.F., Freedman, W.L., Lauer, T., Ajhar, E.A., \& Baum, W.A. 1997b, \apj, 478, 49}
\reference{}{Salaris, M., \& Cassissi, S. 1997, \mnras, 289, 406}
\reference{}{Sasselov, D. \etal 1997, \aap, 324, 471}
\reference{}{Schechter, P.L., Mateo, M., \& Saha, A. 1993, \pasp, 
105, 1342}
\reference{}{Silbermann, N.A. \etal 1996, \apj, 470, 1}
\reference{}{Skillman, E.D., Kennicutt, R.C., \& Hodge, P.W. 1989, \apj, 347, 875}
\reference{}{Skillman, E.D., Kennicutt, R.C., Shields, G.S., \& Zaritsky, D. 1996, \apj, 462, 147}
\reference{}{Soffner, T., Mendez, R.M., Jacoby, G.H., Ciardullo, R., Roth, M.M., \& Kudritzki, R.P. 1996, \aap, 306, 95}
\reference{}{Stanek, K.Z. 1996, \apj, 460, L37}
\reference{}{Stetson, P.B. 1994, \pasp, 106, 250}
\reference{}{Stetson, P.B. \etal 1997, in preparation}
\reference{}{Stift, M.J. 1990, \aap, 229, 143}
\reference{}{Stift, M.J. 1995, \aap, 301, 776}
\reference{}{Stothers, R.B. 1988, \apj, 329, 772}
\reference{}{Webster, B.L., \& Smith, M.G. 1983, \mnras, 204, 743}
\reference{}{Wheeler, J.C., Sneden, C., \& Truran, J.W. 1989, \araa, 27, 279}
\reference{}{Zaritsky, D., Kennicutt, R., \& Huchra, J. 1994,
\apj, 420, 87 (ZKH)}
\end{references}
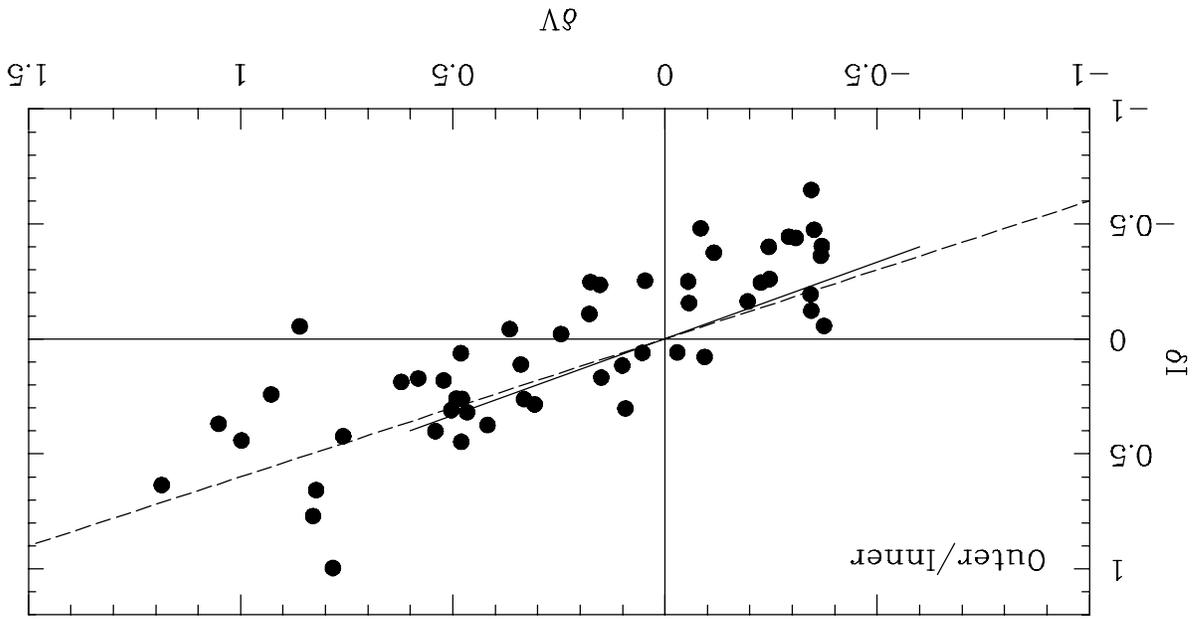
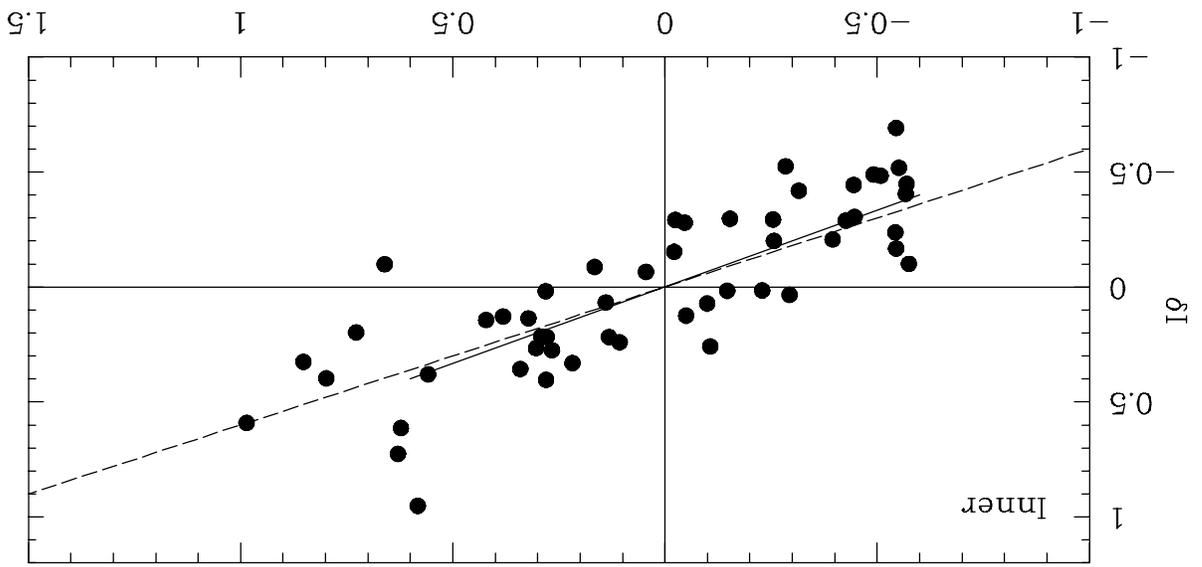
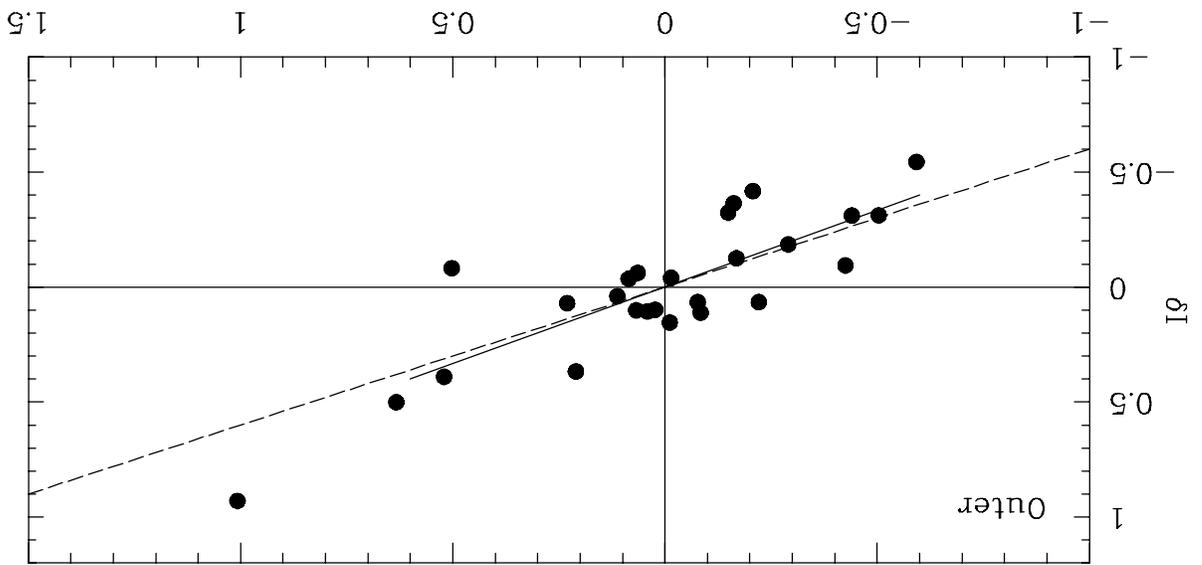
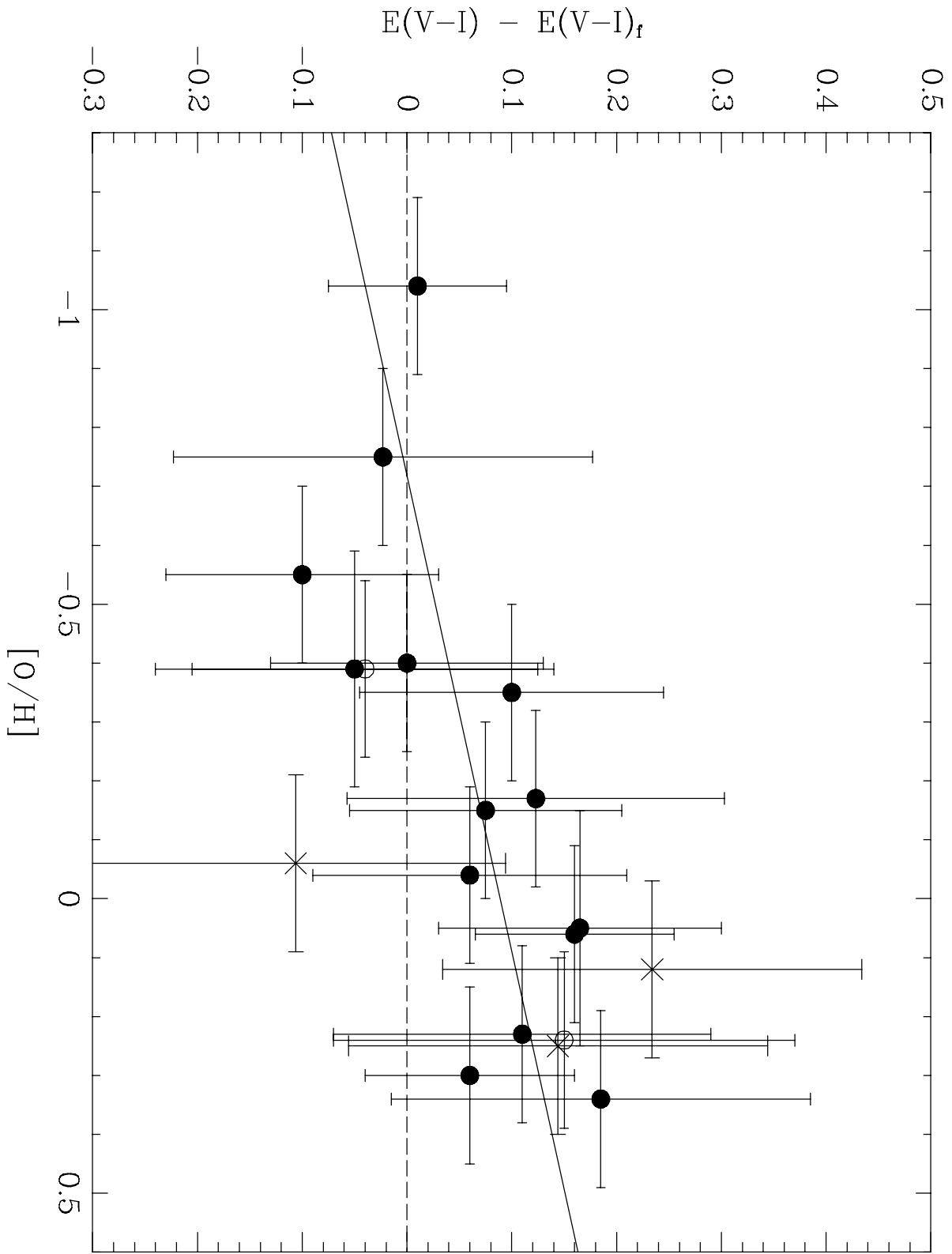
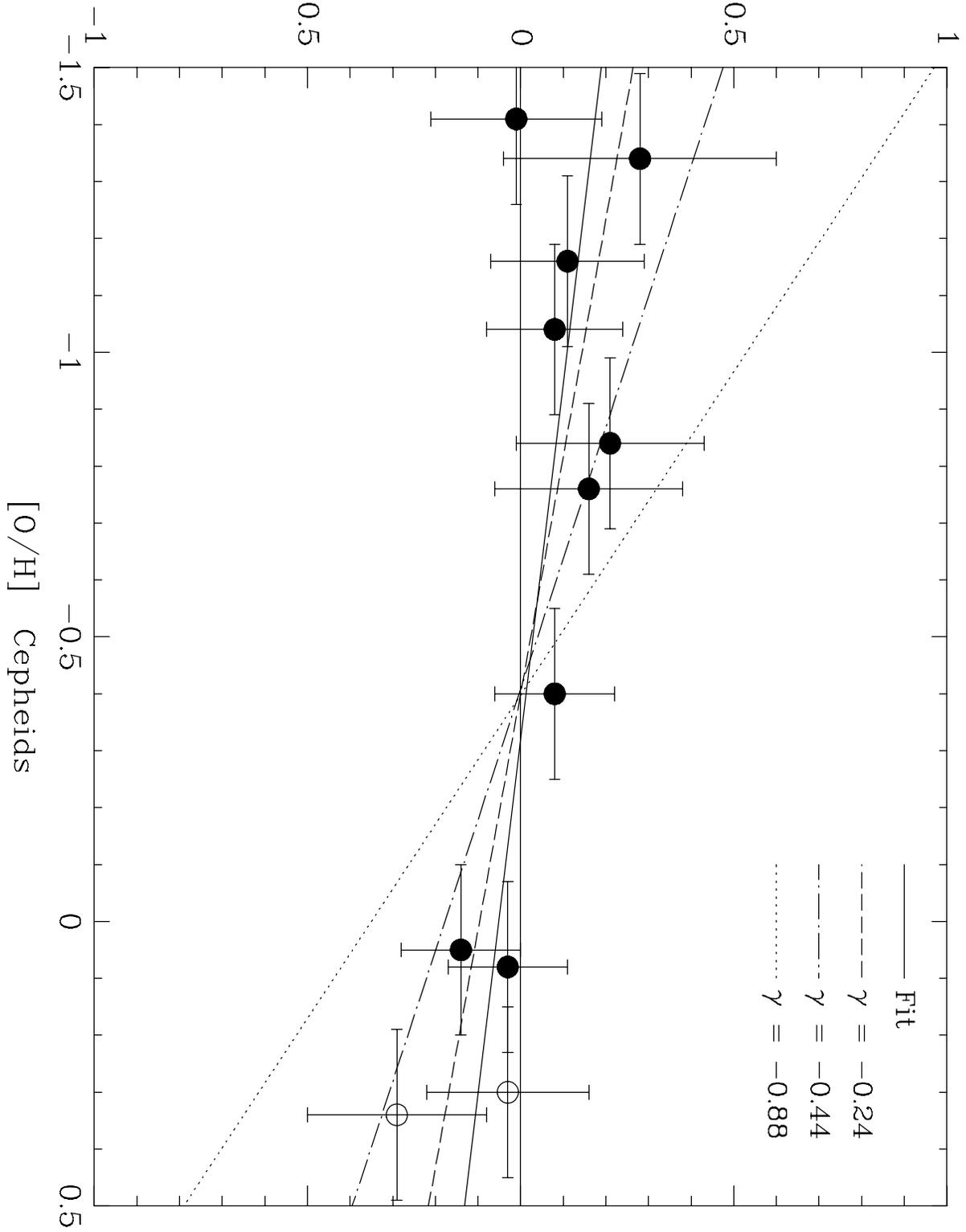
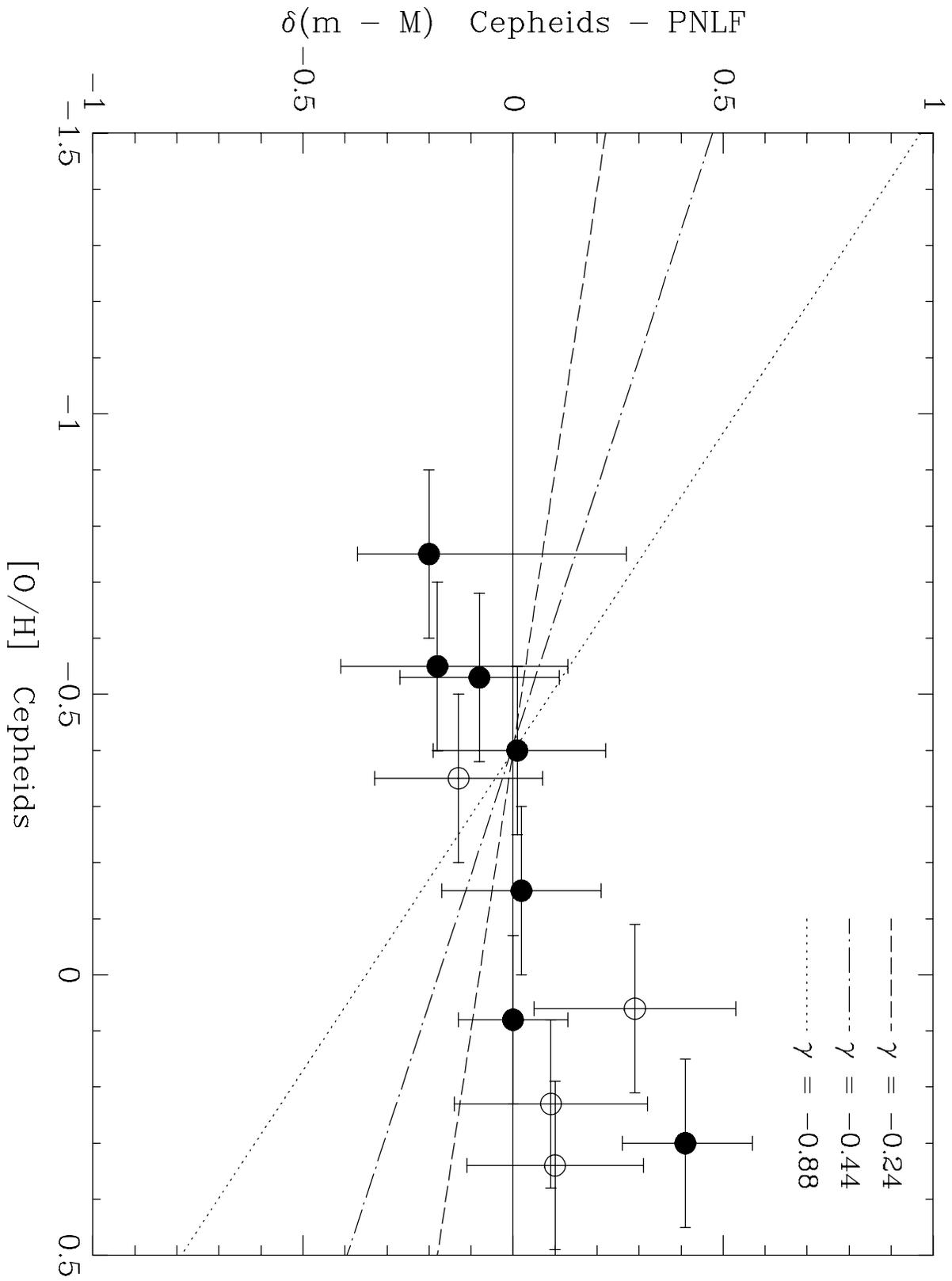
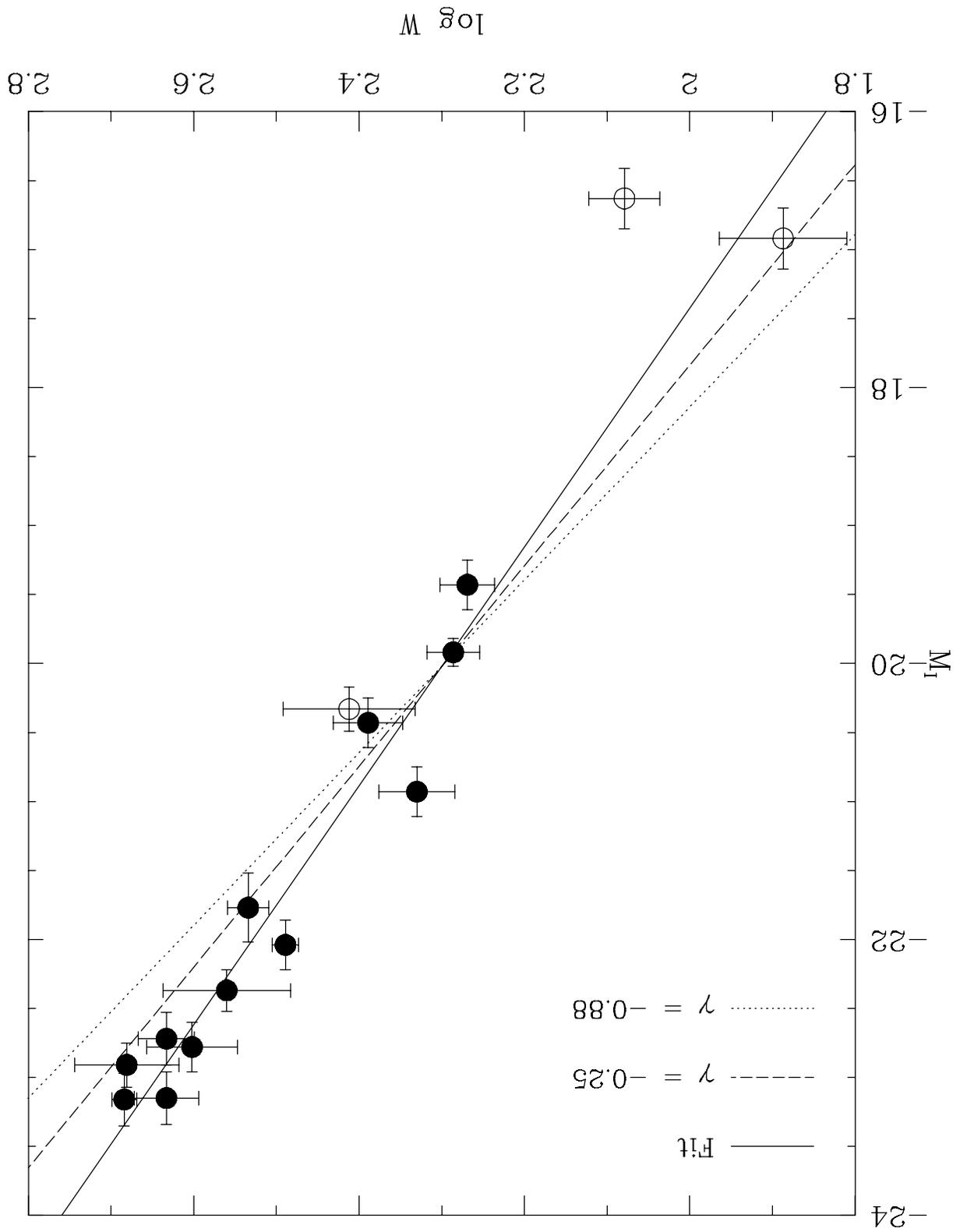
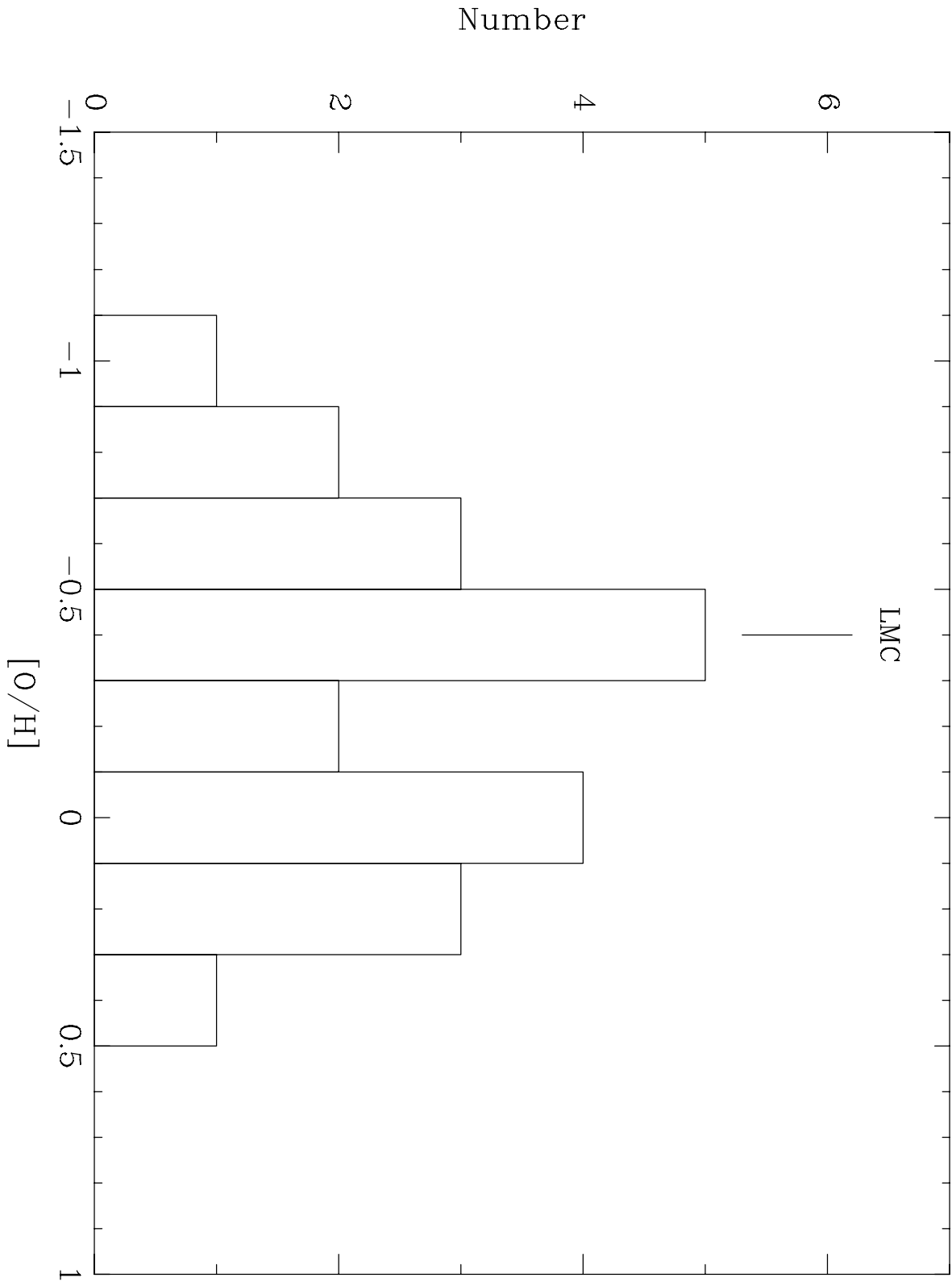
\end{document}